\documentstyle[12pt]{article}

\def\fm3{fm$^3$}
\def\fmm3{fm$^{-3}$}

\begin{document}
\bibliographystyle{unsrt}
%
\begin{center}
{\Large \bf Limits on the neutron--antineutron oscillation \\
time from the stability of nuclei }\\
\vspace{0.4cm}
C.B.~Dover$^{a}$, A.~Gal$^b$ and J.M.~Richard$^c$ \\
\vspace{0.2cm}
{$^a$\ Brookhaven National Laboratory, Upton, NY 11973} \\
{$^b$\ Racah Institute of Physics, The Hebrew University, Jerusalem
91904, Israel} \\
{$^c$\ Universit\'e de Grenoble, Institut des Sciences Nucl\'eaires,
 38026 Grenoble, France} \\
\end{center}
\begin{abstract}
We refute a recent claim by Nazaruk that the limits placed
on the free--space neutron--antineutron oscillation time
$\tau_{{n\bar n}}$ can be improved by many orders of
magnitude with respect to the estimate
$\tau_{{n\bar n}}>2(T_0/\Gamma)^{1/2}$, where $T_0$ is a
measured limit on the annihilation lifetime of a nucleus
and $\Gamma\sim 100$ MeV is a typical antineutron--nucleus
annihilation width.
\end{abstract}
In a recent letter \cite{naz94}, Nazaruk claims to have increased the
limit obtained from the stability of nuclei on the free--space
$n\to\bar n$ oscillation time [2-4] by 31 orders of magnitude.  In view
of the startling nature of this claim, it is important to carefully
inspect the derivation of this result.  In this note, we point
out a specific error in Nazaruk's derivation.  Correcting it, one
obtains a limit of the same order of magnitude as given by previous
authors [2-4] who used potential models for the $n$--nuclear
and $\bar n$--nuclear dynamics.  We also explain in very simple
terms the origin of the standard limit.

\def\nbar{{\bar n}}
Nazaruk's framework is that of the $S$ matrix in the diagrammatic
approach.  For neutron--antineutron oscillations in the nucleus, he
writes down (Eq.~(17) of Ref.~[1]) the probability $W(t)$ for
the transition at time $t$ as
\begin{equation}
W(t) =\epsilon^2t^2-\epsilon^2\int^t_0 dt_\alpha\int^{{t_\alpha}}_0
 dt_\beta W_{{\bar n}}(t_\alpha,t_\beta),
\label{eq:1}
\end{equation}
where $\epsilon=\tau_{{n\bar n}}^{-1}$ characterizes the free--space
oscillation, and $W_{{\bar n}}$ is given by Eq.~(12) of Ref.~[1] (after
noting that $W_\nbar=-2iT_{ii}^{\nbar}$ by comparing Eqs.~(16) and (17)):
\begin{equation}
W_\nbar(t_\alpha,t_\beta)=-2\sum_{k=1}^\infty (-i)^k
\Bigg\langle \int_{{t_\beta}}^{{t_\alpha}} dt_{1\ldots}
  \int_{{t_\beta}}^{t_{k-1}} dt_k H(t_1)\ldots H(t_k)\Bigg\rangle.
\label{eq:2}
\end{equation}
We will follow Ref.~[1] in demonstrating the result for the
oversimplified case where the $\nbar$--nuclear dynamics is reduced to
just a width factor, $H=-i\Gamma/2$, with $\Gamma\sim 100$ MeV being a
typical value.  The introduction of real potentials for the $n$
and $\nbar$ does not change the order of magnitude of the result
\cite{dgr83,alb82}.  The r.h.s. of Eq.~(\ref{eq:2}) is easily evaluated,
resulting in
\begin{equation}
W_\nbar(t_\alpha,t_\beta) = 2(1-\exp[-\Gamma(t_\alpha-t_\beta)/2]),
\label{eq:3}
\end{equation}
which replaces Nazaruk's Eq.~(18)
\begin{equation}
W_\nbar(t_\alpha,t_\beta) = 1-\exp[-\Gamma(t_\alpha-t_\beta)].
\label{eq:4}
\end{equation}
Substituting Eq.~(\ref{eq:3}) in Eq.~(\ref{eq:1}), one obtains
\begin{equation}
W(t)={4\epsilon^2\over\Gamma} t[1-2(1-\exp(-\Gamma t/2))/\Gamma t],
\label{eq:5}
\end{equation}
instead of Nazaruk's erroneous result (Eq.~(19) of Ref.~[1])
\begin{equation}
W(t)=\epsilon^2t^2[{1\over 2}+{1\over\Gamma t} - {1\over\Gamma^2t^2}
 (1-\exp(-\Gamma t))].
\label{eq:6}
\end{equation}
The time dependence of $W(t)$, Eq.~(\ref{eq:5}), becomes linear in $t$
for $t\gg\Gamma^{-1}$, that is for times considerably exceeding the very
short time $\Gamma^{-1}\sim10^{-23}$ sec characteristic of antineutron
annihilation:
\begin{equation}
W(t)\stackrel{\Gamma t\gg 1}{\longrightarrow} {4\epsilon^2\over\Gamma} t =
 {4\over\Gamma \tau_{n\nbar}^2} t.
\label{eq:7}
\end{equation}
Thus the rate of $n\to\nbar$ oscillations in the nucleus is given
by the coefficient of $t$ in Eq.~(\ref{eq:7}).  The nuclear lifetime $T_0$
for annihilation due to $n\to\nbar$ oscillations is given by the
inverse of this rate:
\begin{equation}
T_0={1\over 4} {\Gamma\over\epsilon} \tau_{n\nbar} = {\Gamma\over 4}
\tau_{n\nbar}^2,
\label{eq:8}
\end{equation}
where the first equality shows explicitly the enhancement factor
$\Gamma/\epsilon$ which makes the nuclear lifetime $T_0$ longer by a
huge factor with respect to the free--space lifetime $\tau_{n\nbar}$.
For a measured limit $T_0$ on the stability of nuclei, a limit on
$\tau_{n\nbar}$ emerges:
\begin{equation}
\tau_{n\nbar}>2(T_0/\Gamma)^{1/2}.
\label{eq:9}
\end{equation}
This is the order--of--magnitude estimate, for $\Gamma\sim 100$ MeV, derived
in previous calculations [2-4].

We remark that Nazaruk's
 expression for $W(t)$, Eq.~(\ref{eq:6}), behaves quadratically in $t$ for
$t\gg\Gamma^{-1}$.  This erroneous behavior is due to the overall factor
2 missing in Eq.~(\ref{eq:4}), compared to Eq.~(\ref{eq:3}), so that only
one--half of the free--space $\epsilon^2t^2$ component is cancelled out in
Eq.~(\ref{eq:6}). The full cancellation of the $\epsilon^2t^2$ factor is
equivalent to the statement that disconnected diagrams, such as Fig.~2a of
Ref.~[1], cannot contribute to the $S$ matrix and must therefore be
cancelled by other diagrams (Fig.~2b of Ref.~[1]).  Once the $\epsilon^2t^2$
term of Eq.~(\ref{eq:6}) is dropped, the rest of the terms reproduce
Eq.~(\ref{eq:5}) upon replacing $\Gamma$ by $\Gamma/2$ according to
Eq.~(\ref{eq:3}).

Ref.~[1] also makes strong statements about the inability of (optical)
potential models to produce a correct description of $n\to\nbar$ oscillations
in nuclei.  In fact, once the erroneous result \cite{naz94}  given by
Eq.~(\ref{eq:6}) is replaced by the correct Eq.~(\ref{eq:5}), the potential
model reproduces it exactly.  From Eq.~(22) of Ref.~[1], we have
\begin{equation}
W_{\rm pot}(t)=2{\rm Im} i(\epsilon/\delta U)^2[1-i\delta Ut-
\exp(-i\delta Ut)].
\label{eq:10}
\end{equation}
Substituting $\delta U=-i\Gamma/2$, the potential model expression
$W_{\rm pot}(t)$ is seen to yield precisely the $W(t)$ of Eq.~(\ref{eq:5}),
obtained via the diagrammatic method.

Finally, for clarity, we would like to present a simple derivation of
the $n\to\nbar$ oscillation time in the optical model.  Following
Eqs.~(21) of Ref.~[1], we write the time--dependent coupled Schr\"odinger
equations for a zero momentum neutron and antineutron in external
(nuclear) potentials $U_n=0$ and $U_\nbar=-i\Gamma/2$, respectively:
\begin{equation}
i {\partial\over\partial t} \psi_n=\epsilon\psi_\nbar\quad ,
(i{\partial\over \partial t}+i{\Gamma\over 2})\psi_\nbar=
\epsilon \psi_n.
\label{eq:11}
\end{equation}
Operating on the first equation with
$(i{\partial\over \partial t}+i{\Gamma\over 2})$ and using the
second equation to eliminate $\psi_\nbar$, we obtain
\begin{equation}
({\partial^2\over\partial t^2}+{\Gamma\over 2}\,{\partial\over\partial t}
 + \epsilon^2)\psi_n=0.
\label{eq:12}
\end{equation}
Looking for exponential decay solutions of the form
$\psi_n=\exp(-\gamma t/2)$, where $\gamma$ is the neutron rate of
disappearance, one gets a quadratic equation for $\gamma$
\begin{equation}
\gamma^2-\Gamma\gamma+4\epsilon^2=0,
\label{eq:13}
\end{equation}
with solutions, to leading order in $(\epsilon/\Gamma)^2$, given by
\begin{equation}
\gamma_{_<}\simeq {4\epsilon^2\over\Gamma}\quad , \quad
\gamma_{_>}\simeq\Gamma.
\label{eq:14}
\end{equation}
The solution $\gamma_{_>}$ should be discarded since it corresponds to the
rate of disappearance of an antineutron when its oscillation coupling
$\epsilon$ to the neutron is neglected.  (Note that $\psi_\nbar$ also
satisfies Eq.~(\ref{eq:12}), the difference being in the boundary
conditions.) The rate $\gamma_<$ agrees precisely with that obtained in
Eq.~(\ref{eq:7}), and hence leads to the $n\to\nbar$ nuclear lifetime
expression of Eq.~(\ref{eq:8}).

In conclusion, potential models produce the right temporal evolution of
$n\to\nbar$ disappearance with the correct order of magnitude
estimate for neutron--antineutron oscillation lifetimes in nuclei.
These models \cite{dgr83,alb82}, to various degrees, account also for the
single--particle $n$--nucleus and $\nbar$--nucleus dynamics.
An important observation \cite{dgr83} is that most of the contribution to
$n\to\bar n$ oscillation and subsequent $\bar n$ annihilation in
finite nuclei comes from the outer part of the neutron wave function.
The
inclusion of more sophisticated dynamical effects, such as short range
two--body correlations or medium corrections to the basic process,
is unlikely to change the order of magnitude
of the optical potential results given by Eqs.~(\ref{eq:8}), (\ref{eq:9}).

This work was supported by the U.S. Department of Energy under Contract
DE-AC02-76CH00016. One of us (AG) was partially supported by the Basic
Research Foundation of the Israeli Academy of Sciences and Humanities.

\end{document}